\documentclass{elsart}
\usepackage{graphicx}
\usepackage[footnotesize ,normal, bf]{caption2}
\begin{document}
\begin{frontmatter}
\title{A modular PC based silicon microstrip beam telescope with high speed data acquisition}
\author[Bonn]{J. Treis\thanksref{CORR}},
\author[Bonn]{P. Fischer},
\author[Bonn]{H. Kr\"uger},
\author[Bonn]{L. Klingbeil},
\author[Milano]{T. Lari},
\author[Bonn]{N. Wermes}
\address[Bonn]{Physikalisches Institut der Universit\"at Bonn, Germany}
\address[Milano]{Dipartimento di Fisica di Universit\`a di milano  e INFN, Sezione di Milano, Italy}
\thanks[CORR]{Corresponding author. Tel. +49 228 732499, fax. +49 228 733220, e-mail treis@physik.uni-bonn.de.}
\begin{abstract}
A PC based high speed silicon microstrip beam telescope consisting
of several independent modules is presented. Every module contains
an AC-coupled double sided silicon microstrip sensor and a
complete set of analog and digital signal processing electronics.
A digital bus connects the modules with the DAQ PC. A trigger
logic unit coordinates the operation of all modules of the
telescope. The system architecture allows easy integration of any
kind of device under test into the data acquisition chain.\newline
Signal digitization, pedestal correction, hit detection and zero
suppression are done by hardware inside the modules, so that the
amount of data per event is reduced by a factor of 80 compared to
conventional readout systems. In combination with a two level data
acquisition scheme, this allows event rates up to 7.6
$\mathrm{kHz}$. This is a factor of 40 faster than conventional
VME based beam telescopes while comparable analog performance is
maintained achieving signal to noise ratios of up to 70:1.
\newline The telescope has been tested in the SPS testbeam at CERN. It
has been adopted as the reference instrument for testbeam studies
for the ATLAS pixel detector development.
\end{abstract}
\end{frontmatter}
\section{Introduction}
For the testing of newly developed detector systems, testbeam
facilities are suitable and frequently used. They create
experimental conditions which are closer to a high energy physics
experiment than the conditions in the laboratory while permitting
access to important experimental parameters. In order to measure
properties like efficiency and spatial resolution of a device
under test (DUT), a precise reference measurement of the incident
particle tracks is required. This is the task for a {\em beam
telescope\/} \cite{NIM2,NIM,TEL2} measuring intercept and angle
for incident particles on an event by event basis. In order to
achieve position resolutions in the $\mu \mathrm{m}$ and $\mu
\mathrm{rad}$ scale silicon microstrip detectors are commonly used
for such telescopes, providing a number of space points for track
interpolation. Such microstrip based telescope systems suffer from
limited event rate due to their large number of readout channels
and their system architecture. Additionally, to synchronize such a
system is difficult, and merging a given DUT readout into the
system's data acquisition (DAQ) is a major task.
\newline As beam time is often limited, speed is also an
important requirement for a telescope system, especially when
semiconductor detector devices with a structure size in the
$\mu\mathrm{m}$ scale, for instance ATLAS pixel devices, are to be
tested. The time needed to collect a significant number of events
for every sensor element strongly depends on the readout speed of
the telescope, as the DUT readout is very fast.
\newline In this paper, the concept of a fully PC-based beam
telescope system, henceforth referred to as BAT\footnote{An
acronym for {\em Bonn ATLAS Telescope\/}.}, is presented, which
combines good track measurement accuracy with high event rate and
easy DUT integration.
\begin{figure}
  \centering
  \includegraphics{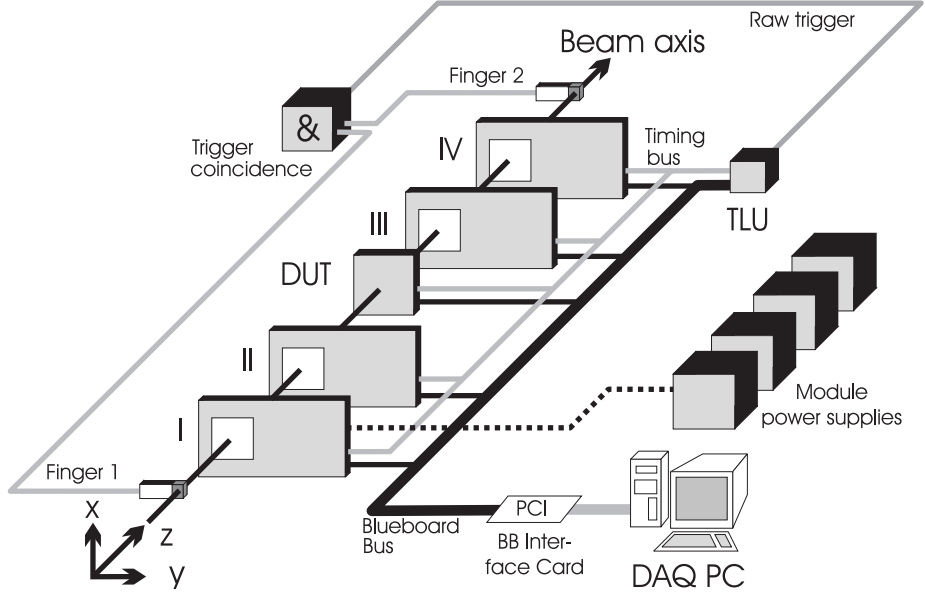}
  \caption{A BAT beam telescope setup with four BAT modules and one
  BB based DUT.}\label{telsetup}
\end{figure}
\section{System overview} Figure \ref{telsetup} shows a typical BAT
setup consisting of four detector modules, a trigger logic unit
(TLU), the data acquisition PC and a DUT. All components are
connected via the purely digital "blueboard bus" (BB) \cite{SISO}.
Furthermore, the "timing bus" connects BAT modules, DUT and
TLU.\newline A raw trigger signal indicating an event is provided
by the coincidence of two scintillation counters. The coincidence
signal is fed into the TLU, which then decides if a trigger is to
be given according to the module's status information accessible
on the timing bus. If so, the TLU generates the trigger signal and
distributes it to the modules.\newline After receiving a trigger,
each module acquires, digitizes and preprocesses event data
autonomously and independent from an external sequencer logic. The
event data is stored in a module-internal memory. When a certain
amount of data is accumulated in a module's memory, the
corresponding module alerts the data acquisition PC to read the
entire data memory content of this module.\newline The DAQ
processes running on the PC collect all data from the different
modules, assemble the data which belong to one event and store it
on the hard disk. Part of the data is processed, the results are
made available to the user for monitoring purposes.\newline
Several ways of integrating a DUT are feasible. The DUT can be
connected directly to the BB, as shown in figure \ref{telsetup}.
For this purpose, a flexible BB interface is available. For
integration of a given VME based DUT and supplementary measurement
equipment, a VME crate can be attached to the DAQ PC using a
commercially available PC to VME interface. And in case an
embedded PC or a VME CPU is to be used for DAQ, a BB to VME
interface has been developed for fully VME based operation of the
entire telescope.
\section{Module hardware}
\begin{figure}
  \centering
  \includegraphics{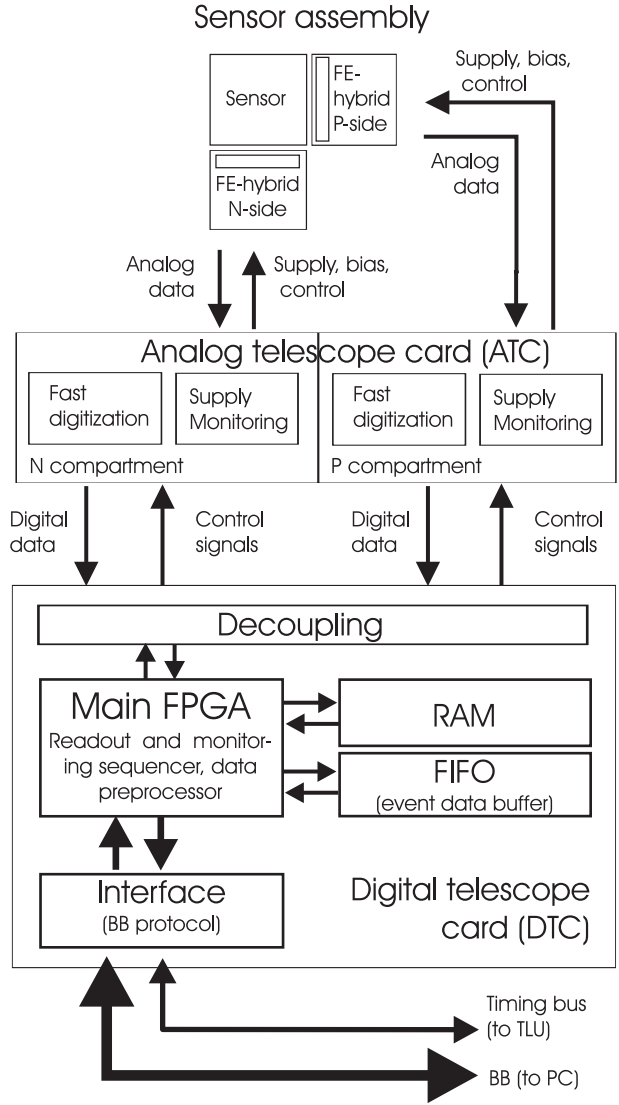}
  \caption{Schematic layout of a BAT module.}\label{modsetup}
\end{figure}

\begin{figure}
  \centering
  \includegraphics{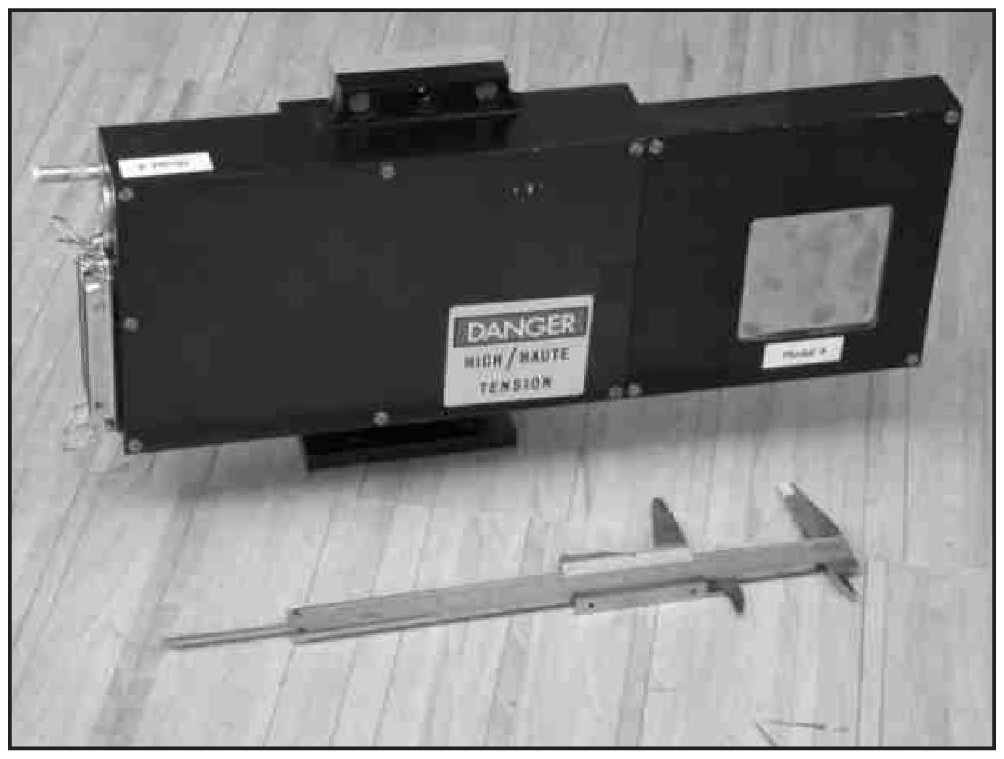}
  \caption{Photo of a BAT module.}\label{modphot}
\end{figure}

\begin{figure}
  \centering
  \includegraphics[width = 250pt]{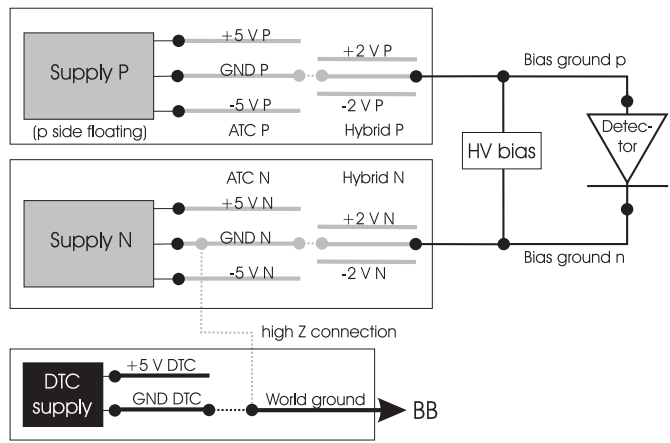}
  \caption{Powering and grounding scheme of a telescope module.}\label{batpotentials}
\end{figure}

A BAT module consists of a sensor assembly, an analog telescope
card (ATC) and a digital telescope card (DTC). An overview over a
module's constituents and their interconnection is given in figure
\ref{modsetup}. A photograph of a fully assembled module is shown
in figure \ref{modphot}.\newline
The sensor assembly consists of the sensor and 2 $\times$ 5 front
end ICs. The sensor is a commercially available double sided,
AC-coupled silicon strip detector type S6934 with integrated
polysilicon bias resistors manufactured by Hamamatsu photonics
\cite{hamam}. The n-side strips are isolated by
$\mathrm{p}^+$-stop implantations. Implant and readout strip pitch
are 50 $\mu\mathrm{m}$ on both sides, the nominal stereo angle is
90$^\circ$. The sensitive area is $3.2\times 3.2\ \mathrm{cm}^2$
corresponding to 640 strips on each side.\newline The front end IC
used is the VA2 manufactured by IDE AS, Oslo \cite{theva}. The VA2
is a 128 channel charge sensitive preamplifier-shaper circuit with
simultaneous sample and hold, serial analog readout and
calibration facilities. Five VA2 ICs are needed to provide readout
for one detector side. They are mounted on a so-called BELLE
hybrid \cite{hybridc,hybridt}, a ceramic carrier with an attached
PCB providing support for the VAs and distributing supply, bias
and digital control to them. As VAs on the hybrid are operated in
a {\em daisy chain\/}, a hybrid is read out like one large 640
channel VA. Sensor and hybrids are fixed to a ceramic support
structure, which is attached to a solid aluminum frame for
handling.\newline The ATC is divided into two identical
compartments supporting one BELLE hybrid each. A hybrid's
compartment provides the supply voltages, bias voltages and bias
currents required by the hybrid. A fast ADC circuit is used for
digitization of the VA2 output data, and an additional
multi-channel ADC allows observation of the most important hybrid
parameters during operation.\newline Only digital signals are
transferred between ATC and DTC via digital couplers. The central
functional building block of the DTC is the {\em readout
sequencer\/}. Implemented into the {\em main FPGA\/}, this circuit
generates the control sequence needed to acquire and digitize the
analog FE data. Both hybrids are read out simultaneously. The
readout sequencer also controls the {\em data preprocessing
logic\/}. Furthermore, the DTC holds a large FIFO for on-module
data buffering and a RAM for storing data preprocessing
information. A second FPGA circuit controls access to the BB and
the timing bus. It is also capable of sending interrupt requests
(IRQs) on the BB to the PC. Each module has its own power supply,
providing three independent voltage sources needed to operate the
ATC compartments and the DTC. The power supply also generates the
detector bias voltage. The powering and grounding scheme of a
telescope module is shown in figure \ref{batpotentials}.
\section{Data acquisition}
\subsection{Structure}
\begin{figure}
  \centering
  \includegraphics[width = 350 pt]{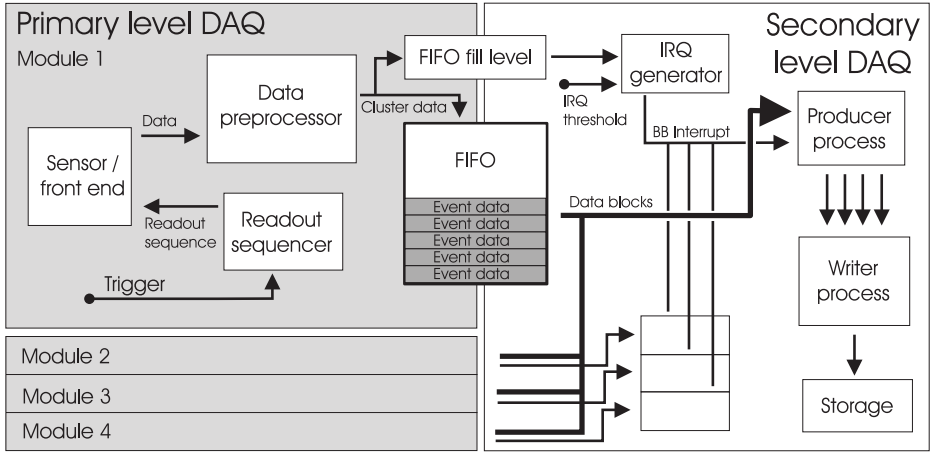}
  \caption{Structure of the telescope DAQ.}\label{daqstruct}
\end{figure}

The data acquisition of the BAT is implemented as a two-level
process. The primary level DAQ (DAQ I), controlled by the readout
sequencers in every module, is simultaneously performed inside
each module directly after receiving a trigger signal. The
secondary level DAQ (DAQ II) is PC controlled and common for all
modules. Both DAQ levels running independently reduces the
effective system dead time to the DAQ I runtime (see also section
\ref{tc}). The telescope DAQ structure is shown in figure
\ref{daqstruct}. An example for DAQ I and DAQ II interaction is
shown in figure \ref{TLUtrig}.
\subsection{Primary level DAQ: Digitizing and preprocessing}
When receiving a trigger, a module's readout sequencer acquires
and digitizes the data residing in the front end ICs and operates
the data preprocessing logic, which performs pedestal correction,
hit detection and zero suppression. Pedestal correction is done by
subtracting an individual pedestal value for each channel. Hits
are detected by applying an individual threshold to the pedestal
corrected data. Pedestal and threshold values have to be
determined and stored beforehand in the DTC RAM. Zero suppression
is done by storing only {\em clusters} consisting of the
information of the 5 neighboring channels around the hit channel
in the DTC FIFO. Enlarged clusters are stored for two or more hit
channels in close proximity. Multiple clusters per event are
possible. The data volume for an event with one hit cluster is 32
byte in total. Compared to a typical event size of 2.5 kByte for
common telescope systems \cite{NIM}, the amount of data is reduced
by a factor 1/80.\newline After finishing preprocessing an event,
end of event data (EOD) is written to the FIFO, which transmits a
module internal trigger number count and the so-called common mode
count (CMC) value. The CMC value is used to calculate and correct
the common mode fluctuation amplitude for this event in on- or
offline analysis. DAQ I has finished processing an event as soon
as a complete module event data block (MED) including cluster data
and EOD has completely been written to the FIFO.
\subsection{Secondary level DAQ: Data readback and event building}
\label{daqq2} While DAQ I is active, MEDs keep accumulating in the
modules' FIFOs until a certain threshold fill level is exceeded. A
module internal interrupt generator generates an IRQ, forcing DAQ
II to become active.\newline DAQ II, responsible for data transfer
to the PC, is controlled by the {\em producer} task, which runs on
the data acquisition PC. It controls one {\em shared buffer}, a
FIFO like structure in PC RAM, for every module. When detecting an
IRQ from a certain module, the producer transfers the data from
this module's buffer FIFO to the corresponding shared buffer. DAQ
I operation is not affected by this data transfer and continues to
process events. The {\em writer} software process collects and
assembles MEDs belonging to the same event from the different
shared buffers and stores them on the hard disk.\newline The
modules' threshold fill level can be adjusted with respect to the
beam intensity. A single event operation mode for low beam
intensities is also available.
\section{Trigger logic}\label{tc} 
\begin{figure}
  \centering
  \includegraphics[width = 300 pt]{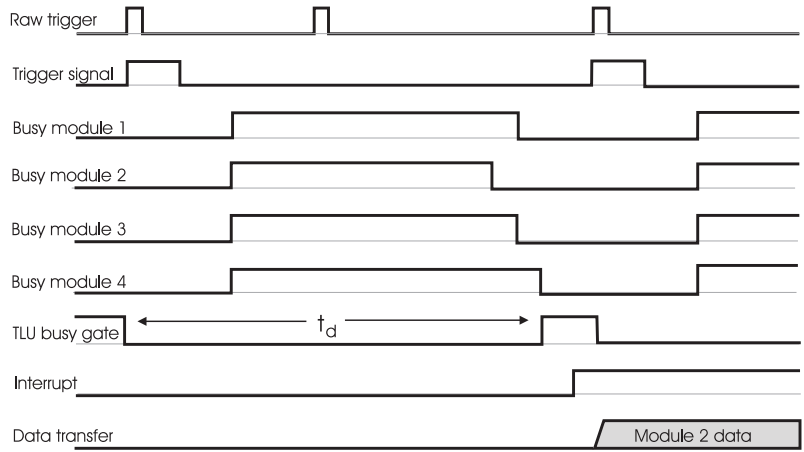}
  \caption{Example for TLU, DAQ I and DAQ II timing. In this example,
  module 2 generates an IRQ right after the first event was fully
  processed. The producer starts the data transfer immediately
  afterwards, while DAQ I processes the second event.}\label{TLUtrig}
\end{figure}

As each module takes data
autonomously, trigger control is necessary to prevent the trigger
synchronization from getting lost. Every device is therefore
connected to the trigger logic implemented in the TLU, receives
its trigger signal from the TLU and has a dedicated busy line on
the timing bus, which indicates DAQ I activity. The TLU generates
a gate signal for the raw trigger from the coincidence of all
devices' busy signals, which only sends triggers if all devices
are not busy. The system's dead time is therefore determined by
the busy signal from the slowest device. The timing of gate and
busy signals is shown in figure \ref{TLUtrig}.
\section{Software} 
\begin{figure}
  \centering
  \includegraphics{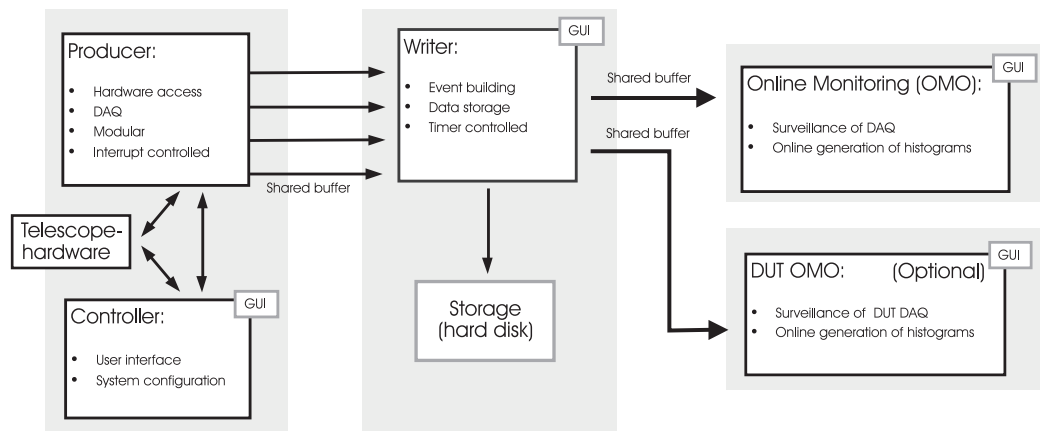}
  \caption{Structure of the BATDAQ software package. The online
  monitoring (OMO) process allows monitoring of telescope module
  data. The DUT OMO is responsible for monitoring of DUT data.}\label{batsoft}
\end{figure}

The DAQ PC is a commercial PC equipped
with a dual Pentium II processor running the Windows NT 4.0
operating system and the DAQ software package written in C++. It
is connected to the BB via a BB to PCI interface card \cite{SISO}.
In addition to the DAQ processes mentioned, online monitoring
processes allow an overview about the device performance during
operation. An overview over the different processes and their
tasks is given in figure \ref{batsoft}.\newline
\section{System performance}
\subsection{Event rate}
\begin{figure}
  \centering
  \includegraphics[width = 250 pt]{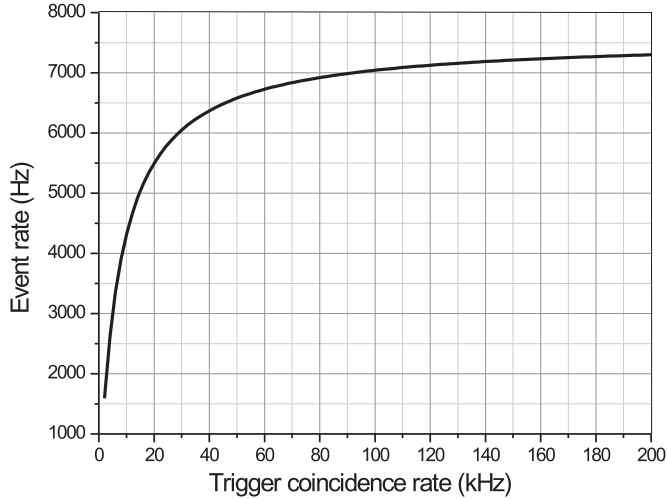}
  \caption{Dependency of event rate on trigger coincidence rate.}\label{trrate}
\end{figure}

The mean event rate of the telescope system is determined by the
dead time of the slowest device, being the BAT modules in most
applications due to their serial readout. A BAT module's dead time
$t_{\mathrm{dt}}$ is dominated by the DAQ I runtime\footnote{The
readout sequencer has to read 640 channels with a serializer clock
frequency of $5\ \mathrm{MHz}$.}, which is 132 $\mu$s. The event
rate actually observed also depends on the trigger coincidence
rate $\Phi_{\mathrm{T}}$, and is given by:
\begin{equation}
{1\over v_{e}^{\, max}} \ = \ t_{dt}\ + \ {1 \over \left(1\, - \,
e^{-\Phi_\mathrm{T}} \right)}
\end{equation}
assuming Poisson statistics. The dependence of event rate and
trigger coincidence rate is shown in figure \ref{trrate}.
At the H8 testbeam at CERN, a system consisting of 4 BAT modules
and one BB based DUT has been operated with an effective event
rate of $4.5\ \mathrm{kHz}$. This is an event rate larger than the
event rate of conventional VME-based systems by factors of 40
\cite {NIM2} to 75 \cite{NIM}.
\subsection{Analog performance}
\begin{figure}
  \centering
  \includegraphics[width = 350 pt]{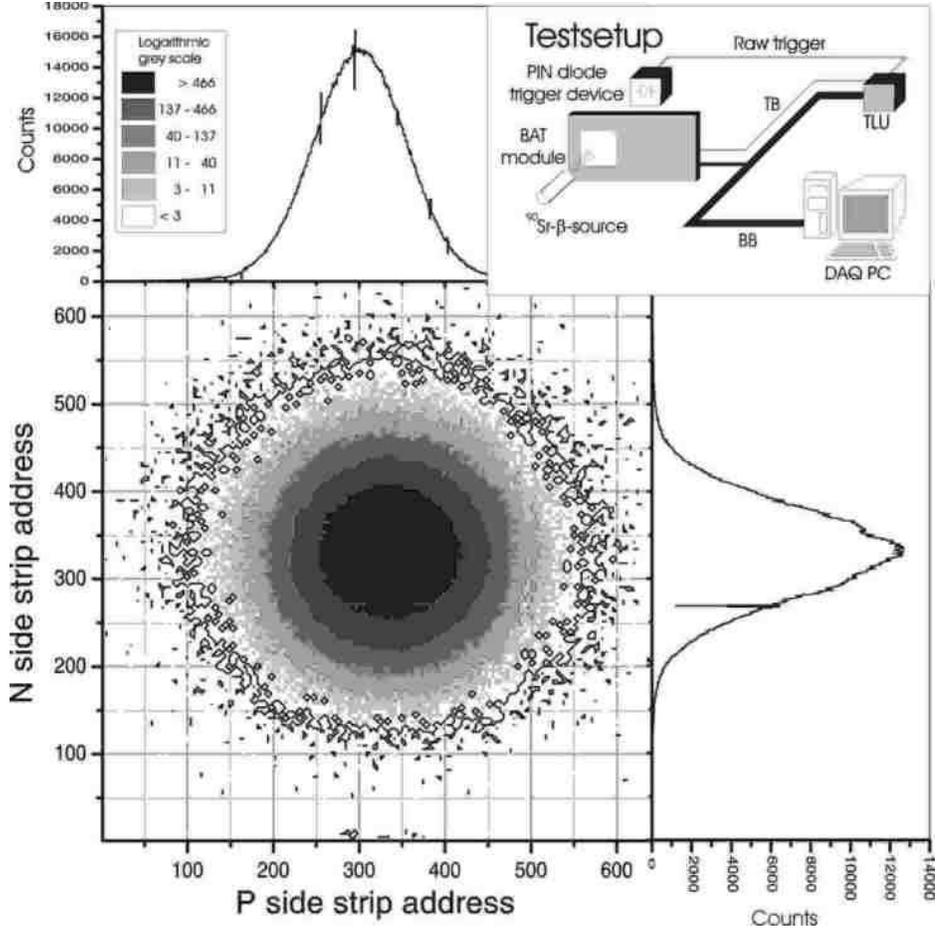}
  \caption{Experimental setup, hit map and source profiles for N- and P- side
  for a source scan of a module using a $^{90}$Sr $\beta$ source with about 2500000 events.
  The number of entries in each bin is grey coded.}\label{hitmap}
\end{figure}

\begin{figure}
  \centering
  \includegraphics{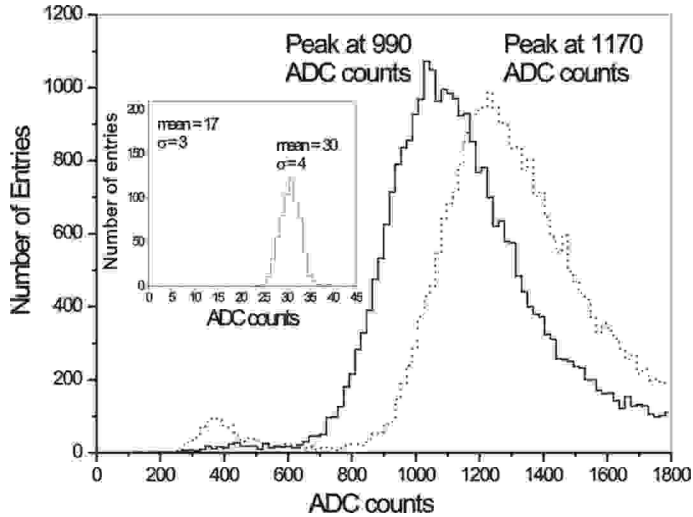}
  \caption{Landau-shaped pulse height distributions of a module in
  a 70 GeV pion beam for N-side (solid line) and P-side (dashed line). The insert
   shows the channel noise histograms for the same module.}\label{perf1}
\end{figure}

\begin{figure}
  \centering
  \includegraphics[width = 300pt]{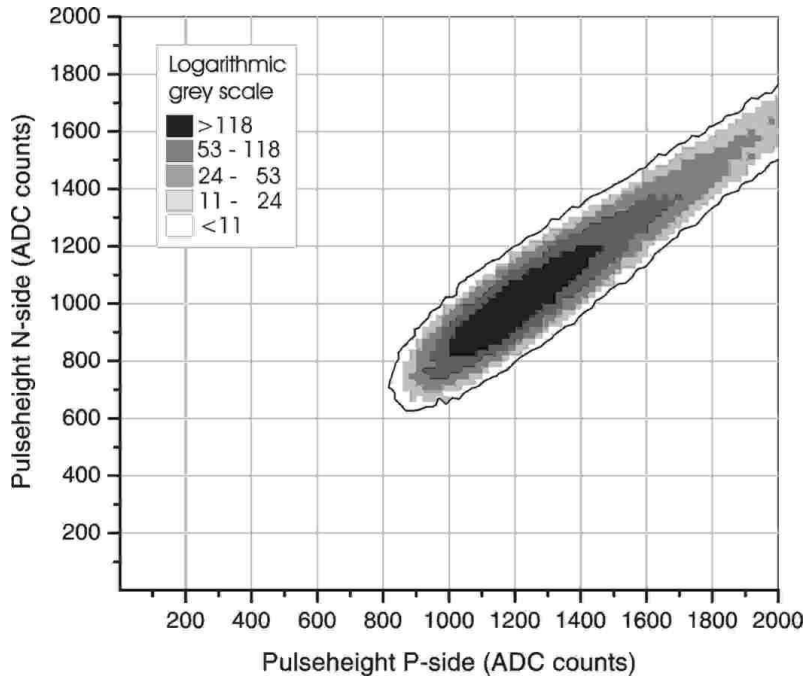}
  \caption{Pulse height correlation between N- and P-side of the same detector module.
  The number of entries in each bin is grey coded.}\label{plscor}
\end{figure}

\begin{figure}
  \centering
  \includegraphics[width = 350pt]{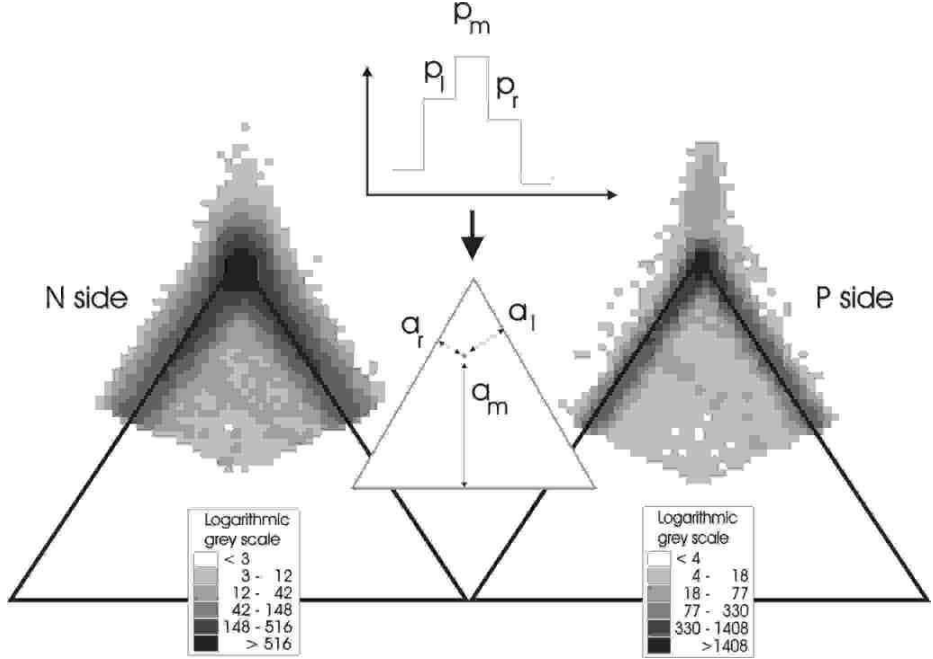}
  \caption{Principle of a triangle plot and triangle plots
  for both detector sides of one module. The $a_l,\, a_m,\, a_r$ are the
  $p_l,\, p_m\, p_r$ values normalized on the sum of the values. The
  number of entries in each bin is grey coded.}\label{dalitz}
\end{figure}

\begin{figure}
  \centering
  \includegraphics[width = 350 pt]{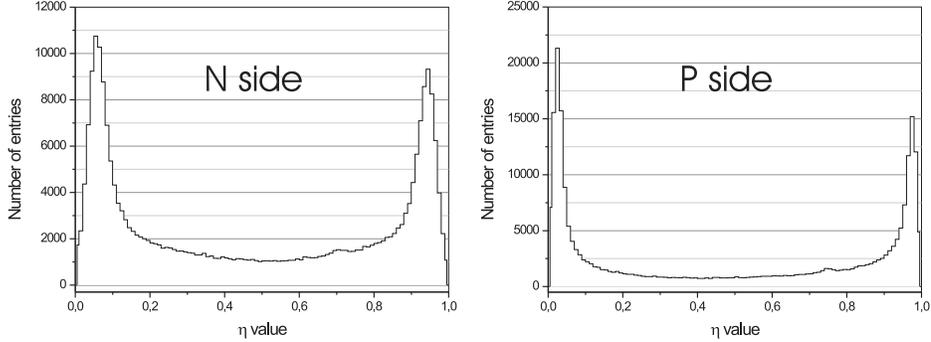}
  \caption{Distributions of $\eta$ for both detector sides.}\label{etadistr}
\end{figure}

Figure \ref{hitmap} shows a hit map and source profiles of a
$^{90}$Sr $\beta$ source scan using a PIN diode as trigger device.
Only one dead channel on the N-side and a few noisy channels on
the P-side are observed. The system operates stably. No pedestal
drift was observed during a 32-hour run. Thus taking pedestals
only once at the beginning of each run is sufficient. Common mode
noise is also tolerable.\newline
Figure \ref{perf1} shows a typical pulse height distribution
together with the noise histogram of N- and P-side of the same
module. One ADC count corresponds to an ENC of 20 $e^-$ on P and
24 $e^-$ on N-side, as can be calculated from the position of the
peak in the respective pulse height distribution. Thus, the mean
ENC value for all channels is 706 $e^-$ for the N-side and 340
$e^-$ for the P-side. Comparing these values with the most
probable charge deposition for a minimal ionizing particle in 300
$\mu$m thick silicon, which is 23300 $e^-$, yields signal to noise
ratios of 33 for the N- and 69 for the P-side, which is comparable
to the results obtained with other telescope systems
\cite{NIM2,NIM,TEL2}.
Figure \ref{plscor} shows the correlation between the pulse
heights observed on N- and P- side of the detector for an event.
The pulse height correlation can be used to solve strip data
ambiguities, which can occur at high beam intensities.\newline
Charge sharing between detector strips can be used for a more
exact reconstruction of the position of a hit on a module, as the
BAT provides analog cluster readout. The normalized pulse heights
of the three central strips of a cluster can conveniently be
displayed in the form of a {\em triangle plot\/} (figure
\ref{dalitz}). Using the different normalized amplitudes of the
three central strips of a cluster as distances from the sides of
an equilateral triangle, the triangle plot is a way to display the
distribution of the signal charge among the three central
channels. Events in which most of the charge is deposited in the
central cluster channel lie at the top of the triangle, events in
which the charge is divided between two cluster channels lie on
the sides of the triangle. Events with significant amounts of
charge on all three central channels lie in the central area of
the triangle.
Entries outside the triangle area are due to "negative" signal
amplitudes after pedestal subtraction caused by noise. In most
cases the charge is deposited only in the central cluster strip or
in two strips. Charge distribution over three or more channels,
mostly due to $\delta$-electrons, are rare; thus an algorithm
using only two cluster charges for reconstruction is appropriate.
The commonly used $\eta$-algorithm \cite{eta} uses the pulse
heights of the two central cluster channels which carry the
largest signals within the cluster:
\begin{equation}
\eta\ = \ {c_1 \over \left(c_1\, + \, c_2 \right)}
\end{equation}
with $c_1,\, c_2$ being the amplitude of the left and the right
central cluster channel. Assuming a uniform distribution of hits
and charge sharing independent from the total pulse height, the
integral of the $\eta$-distribution can be used to calculate a
position correction value $\Delta x$ by
\begin{equation}
\Delta x(\eta) \ = \  {p \over N_0}\int_0^\eta {dN\over d\eta'}
d\eta'
\end{equation}
with $p$ being the strip pitch and $N_0$ the total number of
entries in the $\eta$ distribution histogram. The correction value
is then added to a reference position to obtain the absolute
position of the hit. Typical $\eta$ distributions for a single
module are displayed in figure \ref{etadistr}. The differences in
shape of the $\eta$ distribution between N- and P- side are mostly
due to different interstrip capacitances on the detector sides.
The asymmetry of the distribution for one detector side is due to
parasitic capacitances in the analog readout of the strips. They
can be corrected by applying a deconvolution algorithm. Their
influence on the spatial resolution of the detectors is, however,
small.
\subsection{Spatial resolution}
\begin{figure}[!h]
 \begin{center}
\includegraphics[width=350 pt]{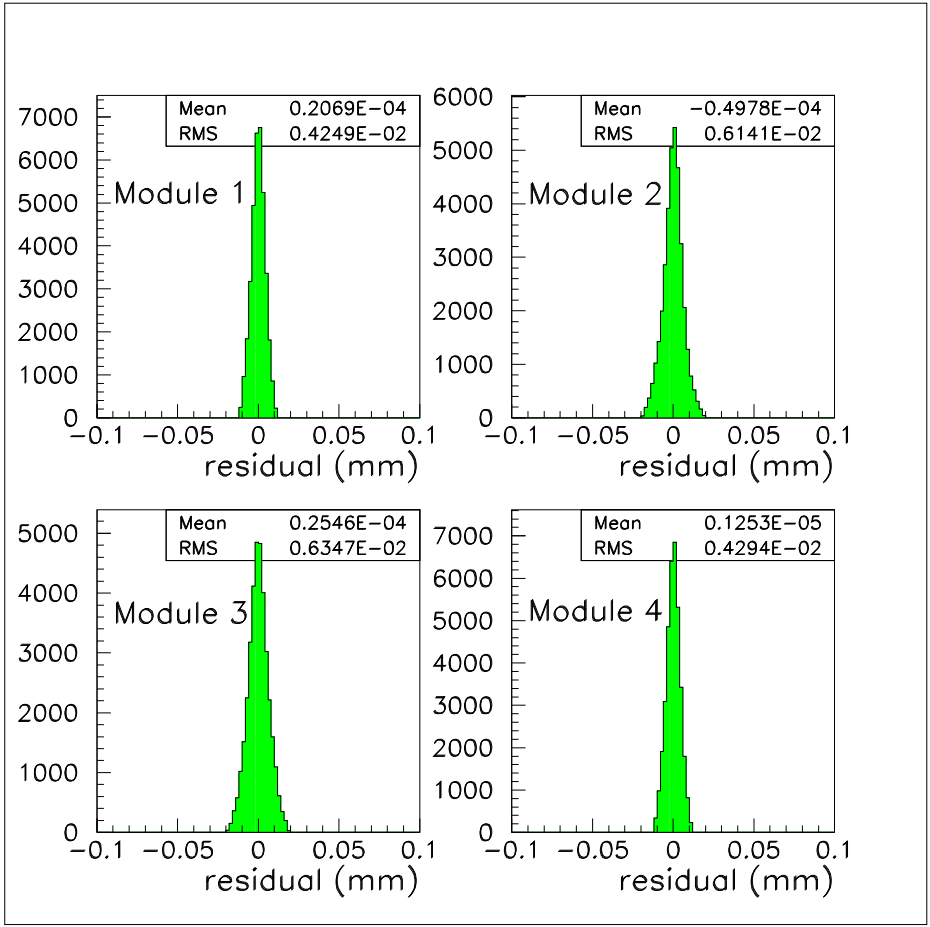}
\caption{\label{StripResiduals} Residual distributions between the
strip hits and the fitted track.}
\end{center}
\end{figure}

\begin{figure}[!h]
 \begin{center}
\includegraphics[width=350 pt]{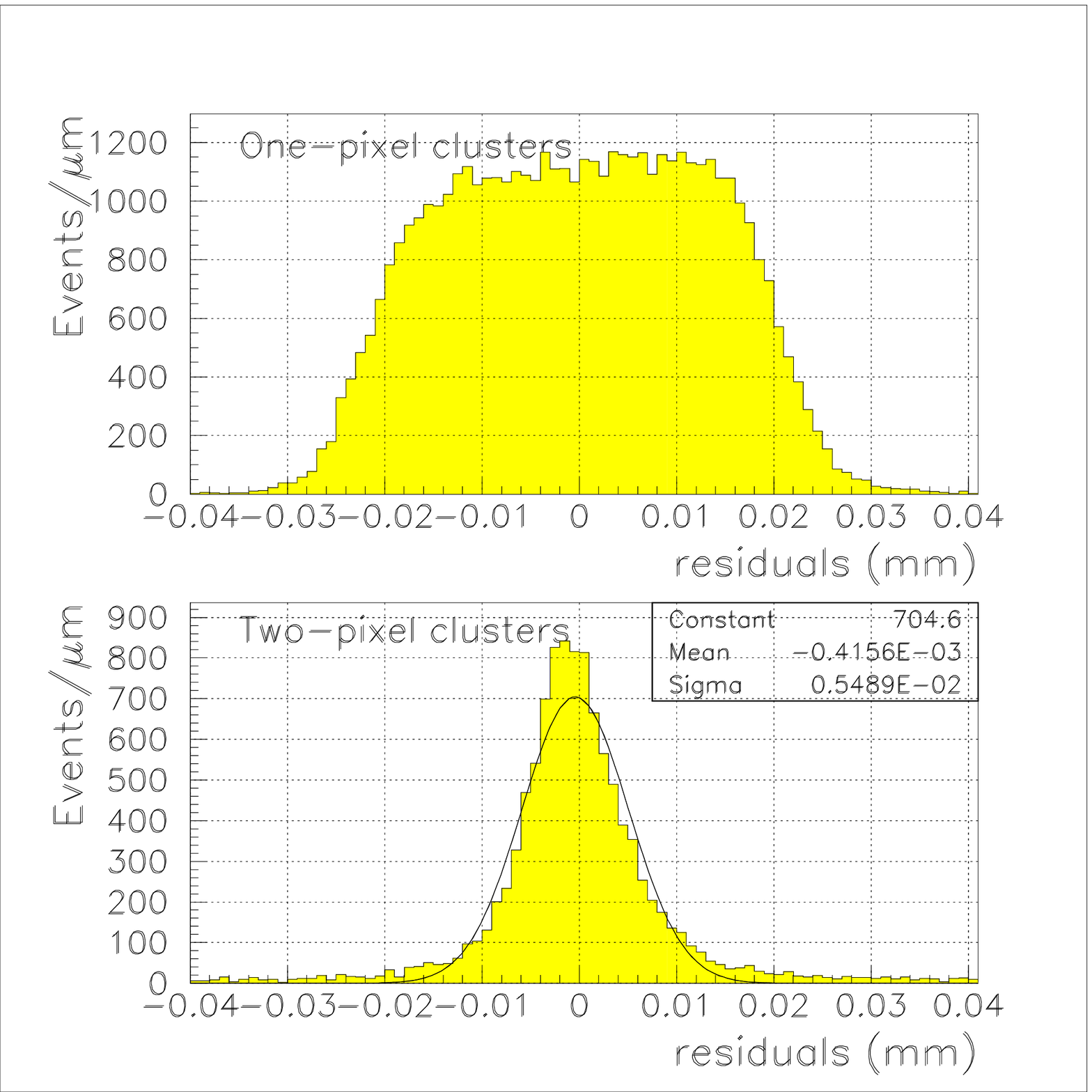}
\caption{\label{PixelResiduals} Residual distributions between the
pixel hits and the extrapolation of the track to the pixel
detector plane along the short direction of the pixel cell. The
upper histogram is for single pixel clusters, the lower histogram
for two pixel clusters with a gaussian fit superimposed.}
\end{center}
\end{figure}

The telescope tracking performance has been studied using test
beam data taken with a 180 GeV/$c$ pion beam at the CERN H8
testbeam at the SPS. The raw event data is processed by a program
developed by the Milano ATLAS group which performs event
reconstruction and alignment of the telescope planes.\newline A
straight line fit is applied to the strip hits, and the residuals
between the hits and the fitted track are computed for the strip
planes. Then, an analytical alignment algorithm is applied to the
strip planes, which minimizes the residuals and their dependence
on position and angle of the tracks. The alignment and the tilt
angle are calculated for all strip planes using the first strip
plane as reference plane. Examples of the resulting residual
distributions for the strip planes in one direction after
alignment are presented in figure \ref{StripResiduals}, showing
the quality of the alignment algorithm. The distributions are
properly centered around zero, which indicates the absence of
systematic errors. Their widths, which are determined by the
intrinsic resolution of the strip planes, multiple scattering and
the alignment algorithm, lie between 6.3 and 4.2 $\mu$m.\newline
As the data from the strip planes, however, is used in the track
fit, the width of the strip plane residuals can not be taken to
determine the spatial resolution of the telescope. For this
purpose, the residual distributions in the DUT planes have to be
considered.
\newline The telescope setup included two DUTs, which were hybrid pixel
detectors. Sensor and front end electronics were developed by the
ATLAS pixel collaboration \cite{talk,sensor}. The sensor has no
inefficient area, the pixel cell size was $50$ $\mu$m $\times$
$400$ $\mu$m corresponding to the pixel pitch, and the front end
electronics provides for zero-suppressed readout, reporting both
pixel position and charge deposition for those pixels only, for
which the charge deposition exceeds a certain threshold.\newline
The spatial resolution of the telescope system was measured using
the residuals between the position determined from the DUT data
and the extrapolation to the DUT plane of the tracks fit to the
strip data. For this purpose, the relative alignment of the DUT to
the strip planes is calculated. Events with a $\chi^2$-probability
of the track fit greater than 0.02 were selected from data taken
with the beam along the normal to the pixel plane. In figure
\ref{PixelResiduals}, the residuals along the short ($50$ $\mu$m)
pixel cell direction are shown for events, for which only one
pixel reported a hit (upper histogram) and events, for which two
neighboring pixels reported a hit (lower histogram). The
reconstructed position on the DUT of the single pixel hits is the
centre of the hit pixel cell, while for two pixel hits an
interpolation algorithm is used to determine the hit position
using the charge deposition information \cite{LorentzPaper,tesi}.
The latter distribution can be used to give an estimation of the
telescope resolution.
\newline
A gaussian fit to the two pixel residual distribution yields
$\sigma = 5.5 \ \mu$m. This is the convolution of the telescope
resolution and the pixel detector intrinsic resolution. The latter
can be estimated as follows. As tracks are uniformly distributed,
the width $L$ of the region in which charge division between two
pixels occurs can be estimated using the ratio between the number
of two pixel and one-pixel hits. This yields $L \simeq 10$ $\mu$m.
The expected r.m.s. of the residual distribution for these tracks
is $\sigma = L/\sqrt{12} = 2.9$ $\mu$m. Thus, the width of the
actual residual distribution is dominated by the telescope
resolution, which can be estimated conservatively to be better
than $\sigma \, = \, 5.5 \ \mu$m in the DUT plane.
\section{Summary}
A high speed modular PC based beam telescope using double sided
silicon microstrip detectors with on module data preprocessing has
been built and successfully taken into operation. Telescope hard-
and software are capable of stand-alone operation and easy to
handle; integration of an additional "device under test" is
straightforward. Pedestal subtraction, hit detection and zero
suppression are done inside every module, reducing the data volume
by a factor of 1/80. With its two level data acquisition scheme,
the system can process event rates up to 7.6 kHz. The telescope is
a factor of 75 (40) \cite{NIM} (\cite{NIM2}) faster than
conventional VME based systems while providing comparable
performance. Signal to noise ratios of up to 70 were achieved. The
spatial resolution in the DUT plane has been determined to be
better than $5.5\ \mu$m.
\begin{ack} We gratefully acknowledge the help obtained from
Walter Ockenfels and Ogmundur Runolfsson when encountering
problems concerning mechanics, case design and handling and
bonding of silicon detectors. We would also like to thank the
members of the ATLAS pixel collaboration, in particular John
Richardson from LBNL, Berkeley, and Attilio Andreazza, Francesco
Ragusa and Clara Troncon from the Milano ATLAS group, for
providing help and know-how in testbeam data taking and data
analysis.\newline
\end{ack}
% ----------------------------------------------------------------

\end{document}